\documentstyle[prl,epsfig,twocolumn,aps]{revtex}
\begin{document}
\twocolumn[\hsize\textwidth\columnwidth\hsize\csname@twocolumnfalse%
\endcsname
\title{Thermal noise and the branching threshold in brittle fracture}
\author{L. M. Sander$^1$ and S.V. Ghaisas$^2$} 
\address{$^1$
Physics Department, The University of Michigan, Ann Arbor MI 48105-1120. USA\\
$^2$ Department of Electronics Science, University of Pune, Pune 411007, India}

\date{\today}

\maketitle 

\begin{abstract} 
Many studies have confirmed that cracks in brittle materials branch
when the crack speed exceeds a certain threshold, $v_c$, but the value
of that threshold is not understood.  Almost all theoretical
calculations overestimate $v_c$ by factors of two or more.  We show
that thermal noise can reduce the threshold by a substantial amount,
and we propose that this effect can account for the discrepancy.

\end{abstract} 

] 
\narrowtext
The formation of cracks in the fracture of solids leads to extremely
intricate pattern formation.  One of the most
interesting phenomena in this field is the mirror-mist-hackle
transition in brittle fracture, for example, that of glass \cite{Lawn}.
This occurs when a crack accelerates and passes through  
a critical velocity, $v_c$ above which
it becomes unstable against branching \cite{bc2,sgf6,sgf5,bcs}.
Below $v_c$ the surface created by the crack is smooth  (mirror-like)
while above it is rough.  However, the value of $v_c$ is not
theoretically understood: typically, theoretical estimates are too
large \cite{lm,ffa}. In this paper we will show that including thermal noise in
the theory can lower the threshold substantially.

For many conditions, once started, cracks accelerate until
their velocity is of the order of the free surface wave (Rayleigh)
velocity $c_{R}$. In the case of a Mode I crack, the
crack cannot move faster than the sides can move apart: that is,
faster than $c_{R}$.  Thus $c_{R}$ is an upper bound on the crack
speed \cite{fre}. However, in practice, cracks travel slower than $c_{R}$.  
This has been associated with an instability against branching since the
seminal work of Yoffe \cite{yo}.  She showed analytically that a
straight crack is intrinsically unstable at speeds above $\approx 0.66
c_{R}$.  The idea is roughly as follows: at low speeds 
there is stress concentration in front of a crack leading
to further motion.  However, at high speeds the stress pattern becomes 
more isotropic, and above a velocity
threshold the maximum stresses are at a finite angle to the direction
of the tip; this may be assumed to give rise to branching.  The effect
is due to the fact that the stresses are transmitted by
sound waves: if the speed of the crack is too large the crack 'catches
up' with the stresses.  Other views of the branching instability have
been formulated  \cite{ben} which differ in detail, but not in
principle, from Yoffe's work.  Simulations of 2-dimensional 
spring models \cite{lm,ffa} also give a branching instability around
0.66$c_{R}$, and the model we use here has the same property.  
However, the experiments always show branching at still
lower speeds $v_c \approx 0.36c_R$ in the case of PMMA, a nearly ideal
brittle material; similar observations
have been made in ordinary glass  \cite{bc2,sgf6}. It
appears that something is causing branching before the Yoffe mechanism
takes hold. Here we introduce a physical effect that has been
left out of previous treatments, thermal noise, and show that it
could resolve the discrepancy.  

It is obvious that large enough thermal noise could cause a 
crack to wander.  The essential
question is whether reasonable values for the parameters give rise to a
substantial effect.  {\it A priori} this seems unlikely:  
usually, bonds break due to thermal effects only near the
melting point, which is not the situation at hand.  However, this is
not correct because the nature of the branching
is that that even below the threshold the stress pattern
near the crack tip is approaching instability, so that a relatively small
thermal fluctuation could break the 'wrong' bond, and give rise to
branching. We will demonstrate this effect in a simulation where we
attempt to fit {\it all} of the parameters to experimental values.

To account for thermal noise, we need to have a coupling between the
long-wavelength elastic modes, which we describe by a spring model,
and a 'heat bath', {\it i.e.} the rest of the degrees of freedom of
the solid. This arises from a damping or viscosity term.
In the macroscopic theory of elasticity
damping can be included in the elastic equations of motion
by introducing Kelvin viscosity terms. With these terms
the equations of motion for the spring model we use \cite{pgl,pgl2} are: 
\begin{eqnarray}
m  \partial_{tt} {\bf u}_{\bf r}  &=&  \sum_{\bf r'} K
{\bf \hat n} \left[{\bf \hat n}\cdot
( {\bf u}_{\bf r} - {\bf u}_{\bf r'} ) \right] \nonumber \\ 
&+& \sum_{\bf r'} \eta
{\bf \hat n} \left[{\bf \hat n}\cdot
\left(\partial_t {\bf u}_{\bf r} 
-\partial_t {\bf u}_{\bf r'} \right) \right]
\label{eqmotion}
\end{eqnarray}
where {\bf u$_{\bf r}$} is the displacement of the node at ${\bf r}$ , $K$ and 
$\eta$ are the spring constants and viscosity respectively, ${\bf
{\hat n}}$ is the unit vector from node ${\bf r}$ to ${\bf r'}$,
and the ${\bf r'}$'s are the nearest neighbors of ${\bf r}$. 
The viscous terms are the simplest form consistent
with the symmetry of the situation \cite{Landau}. We have previously
shown that if $\eta$ is of order unity
this term is sufficient to allow a crack to propagate
without branching at a {\it terminal velocity} less than $v_c$.
Without viscosity any crack, once initiated in in fixed
grip loading conditions (which is the case we consider), 
accelerates to the branching threshold \cite{pgl,pgl2}.

The value of $\eta$ is central to our theory.
We get it by fitting to the experimental values for the attenuation 
of longitudinal sound \cite{maris}. It is
easy to show from Eq. (\ref{eqmotion}) that $\alpha(\omega)$,
the inverse of the attentuation length for longitudinal sound
of frequency $\omega$, is given by $(\eta/2Kc_L)\omega^2$ in our spring model.
We define a characteristic frequency, 
$\omega_o = c_L /a$ for any elastic system, where $c_L$ the 
speed of longitudinal sound, and $a$ is the length
unit, which we take to be the radius of the crack tip. 
(It is well known that any fracture model
must introduce some microscopic length unit.) Then we put 
$\alpha a = Q (\omega/\omega_o)^2$. The dimensionless
number $Q$ is a measure of the size of the damping. For the
spring model, in units where $K=m=a=1$ we find that $Q=0.53 \eta$.
For SiO$_2$ the sound attenuation has also goes like $\omega^2$ over a
large frequency range \cite{maris}. Using the measured data in Ref. \cite{maris}
for $\alpha$ and $c_L$ and using $a \approx 0.5$ nm, a molecular dimension,
and we find $Q \approx 0.4$.
For the case of PMMA the attenuation does {\it not}
have the same frequency dependence
\cite{mor}, and we cannot unambiguously determine $\eta$
\cite{pgl,pgl2}. In the remainder of this paper we will 
mostly discuss SiO$_2$. For this
case we are justified taking $\eta$ to be of order unity. In all of our
simulations we take $\eta=2$. 

In order to simulate fracture we take Eq. (\ref{eqmotion})
as the dynamics of a snapping spring model:
If any bond is stretched more than $u_{th}=0.1a$ we break it. Strips of lattice 
along the $x$ direction are
stressed by holding two edges parallel to the $y$-axis at fixed
displacement, $\Delta$, and a crack is started parallel to the $x$-axis. 
We artificially
cut the springs at an effective crack velocity of 0.05 \cite {pgl,pgl2} at
the center of the width along the length. We solve the equations
of motion by using a Verlet algorithm.
As soon as velocity exceeds
0.1 the artificial scissoring is stopped. The results are
unaffected by this procedure. 
The width ($y$-direction) along which the strain is applied is up to $100a$
and the length ($x$-direction), along which the crack propagates
is up to $400a$.  For
more details of our procedure see \cite{pgl,pgl2}.

\begin{figure} 
\epsfxsize=\hsize \epsfysize = 3.5in \centerline{\epsfbox{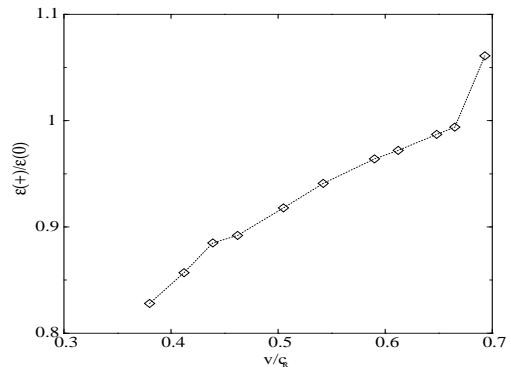}}
\caption{The branching instability as a function of $v$. The strain
in the row corresponding to forward growth is compared to that in
the next row.}\label{yofffig}  
\end{figure}

We first show that the Yoffe instability, or something very much like
it, is in fact present in our model. We create cracks at various terminal
velocities by varying $\Delta$, and measure the ratio of the
strain in the 'right' bond
(corresponding to a straight crack) and the 'wrong' one. This is shown
in Fig. (\ref{yofffig}). The ratio goes through unity and the
crack branches at $\approx 0.7  c_{R}$.

Now we add thermal noise by putting a random force term, ${\bf\theta}({\bf r},t)$,
 on the right hand side of Eq. (\ref{eqmotion}).
The components of the force are independent gaussian noises 
at each site whose amplitude is chosen to be consistent with
the fluctuation-dissipation theorem:
\begin{equation}
<\theta_i({\bf r},t)\theta_j({\bf r'},t')>
 =\frac{2k_{B}T\eta}{dt}\delta_{t,t'}\delta_{r,r'}\delta_{i,j}
\label{noiseeq}
\end{equation}
where $i, j$ label the components.
The dependence on $1/dt$ is the standard way to normalize a discrete 
$\delta$-function.

Strictly speaking the noise should have nearest
neighbor correlations in order to be consistent with our form of
$\eta$. Put another way,
short wavelength modes have more damping than long wavelengths 
as mentioned above, and we should drive them more. Since bonds are
broken by short wavelength fluctuations, we are {\it underestimating}
the effect of noise by neglecting the correlations.
We have done a few runs with the completely 
consistent noise to verify this. The difference is not very
large.      
  
We will need to know the energy unit for our model, $Ka^2$,
in order to evaluate the noise strength. This we get by
fitting $Ku_{th}^2/2a^2$ to the fracture energy, $8 J/m^2$
for SiO$_2$ \cite{Lawn}. Taken literally, this would give each
bond about 20 eV, which is too large by a factor of 10 or so.
In our units room temperature comes out to be of order $10^{-5} -
10^{-6}$. We put $k_{B}T=3 \cdot 10^{-6}$. By taking
a large energy unit we are once more underestimating the noise 
strength.  

In the vicinity of the tip there are large
local heating effects  in some materials including PMMA \cite{f3}.
To account for this we start the lattice with an ambient temperature
(room temperature) and allow
it to locally heat (and diffuse heat) from the work of
the viscous forces. In that case the temperature becomes a
local variable $T({\bf r})$. In order to find $T({\bf r})$ we
set:
\begin{equation}
\partial_t T({\bf r}) = \frac{1}{C} [({\bf f}_{v}({\bf r})-
{\bf \theta}({\bf r})]\cdot\partial_t {\bf u}_{\bf r}
+ \kappa \nabla^2 T({\bf r}) 
\end{equation}
Here $\kappa$ is a thermal conductivity, $C$ is the heat
capacity associated with the local heat bath, and ${\bf f}_{v}$
is the viscous force on node ${\bf r}$ (the second term
on the right-hand side of Eq. (\ref{eqmotion}). If $C$ is very
large the lattice  remains at the ambient temperature. 
In our simulations we adjust $C$ and $\kappa$ so that the vicinity of the tip
heats up to about three times room temperature. 

The simulation precedure is similar to that in
the noise-free case outlined above. We use a version
of the Verlet algorithm modified to account for the random forces
$\theta_i({\bf r},t)$. An initial crack of 
length $50a$ is created
and the lattice is allowed to thermalize. After the temperature 
has settled down,
the lattice is strained to the desired value. 

\begin{figure} 
\epsfxsize=\hsize \epsfysize = 3in \centerline{\epsfbox{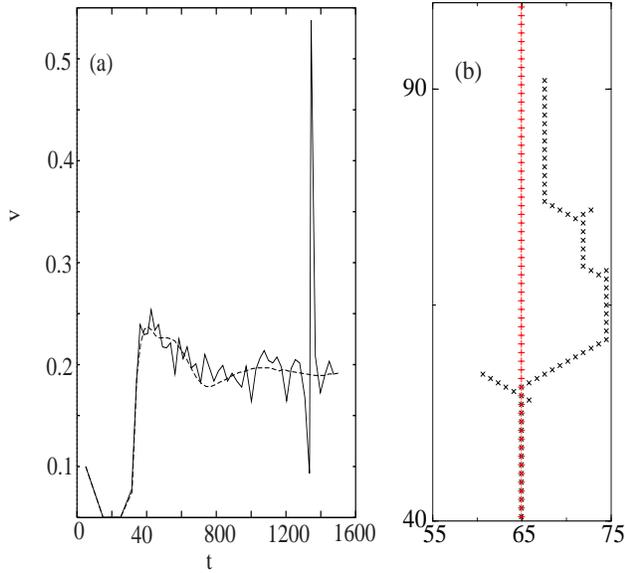}}
\caption{Effect of noise on  crack propagation. (a.) Dashed line is 
the velocity for a 50 x 400 lattice for T=0$^0$K, $\eta$=2.0, 
specific heat $C=100$, $\kappa$=4.0, and stretch per unit 
width $\Delta/W$=0.033 where $W$ is the lattice
width. The solid line is the velocity for the 
same parameters with $T=3.0 \cdot 10^{-6}$ (b.) Crack pattern with and
without noise in the same loading conditions}\label{velnoise}  
\end{figure}

Our results are as follows. Since most of the detailed observations
in the literature are for PMMA, we will compare to the qualitative
features seen in that material, even though the quantitative fitting
that we have done is for glass.
Fig. (\ref{velnoise}a) shows the effect of thermal noise on the crack velocity.
Noise leads to fluctuations in the crack velocity, an effect
that is experimentally observed \cite{bcs,sgf6}. The large change
in the velocity in the figure marks the change in the row of propagation.   
As is seen from the figure, 
there is small rise in the velocity before it saturates. This is also observed 
when the fracture energy is low \cite{lm}. In Fig. (\ref{velnoise}b) we show two
cracks, one without noise and one with, in otherwise identical conditions. 

\begin{figure} 
\epsfxsize=\hsize \epsfysize = 3in \centerline{\epsfbox{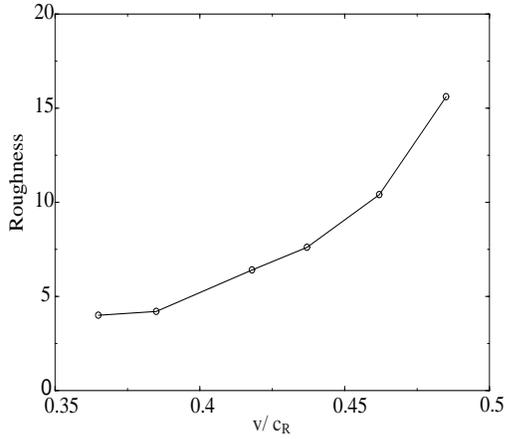}}
\caption{Rms deviations in the crack path as a function of velocity 
for  $\eta=2.0, C=100, \kappa=4.0, T=3.0 \cdot 10^{-6}$.
$\Delta$ is varied to  obtain the
appropriate velocity.}\label{noisevc}  
\end{figure}

With noise there is a finite probability of wandering for any $v$. Thus we
must define the critical velocity in terms of the roughness of the 
crack. We call $v_c$ the velocity above which the roughness
is  more than 10\% of the width. 
Beyond this point the fluctuations grow quickly and saturate due to the 
finite size of the sample. This observation is consistent with the experiments \cite{bcs}.   
Fig. (\ref{noisevc}) shows the effect of thermal noise on $v_{c}$. Here the rms deviations 
are measured from the 
change of direction  and are averaged over 5 runs for each point.    
The critical velocity for crack wandering is lower than in the 
absence of noise and is around 0.45$c_{R}$. 
  
\begin{figure} 
\epsfxsize=\hsize \epsfysize = 4in \centerline{\epsfbox{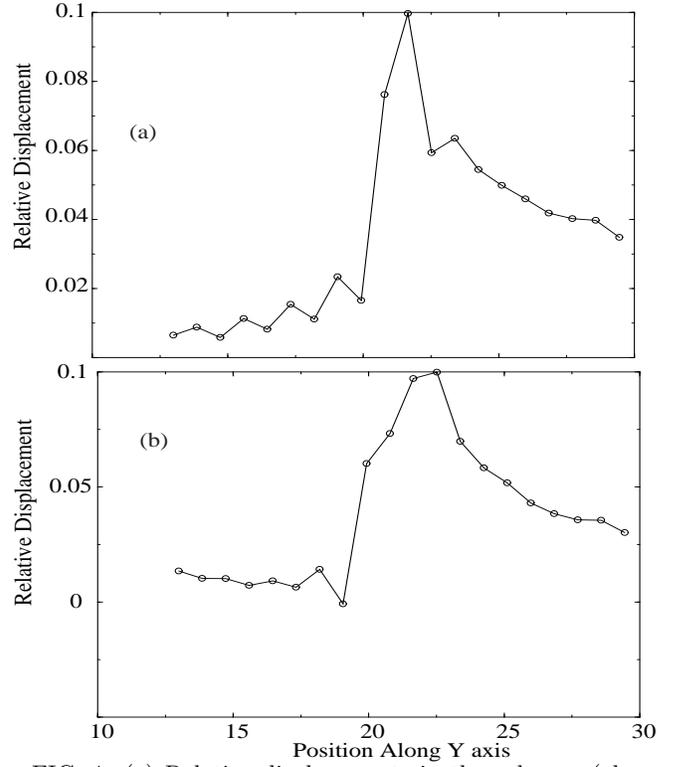}}
\caption{(a) Relative displacements in the columns (along $y$ direction)  
containing the crack tip and one column ahead  
just before the crack begins to propagate. (b) Relative displacements 
in the columns containing the tip just before first wandering is encountered.
The parameters are the same as in the previous figures.}\label{wander}  
\end{figure}

The effect of crack wandering from one plane to another causes the 
rough interface. The experiments \cite {bcs} show that at higher velocities 
the roughness of the cracked surface is larger.
Similar effects occur \cite{hm} for simulations 
of crack propagation in silicon.  
The nature of the wandering instability on our simulations
is interesting. Just above $v_c$ the crack does not change
its direction of propagation.  In our simulations this occurs
for velocities higher than 0.7$c_{R}$. 
At lower $v$ the crack wanders by {\it jumping rows} 
to planes adjacent to the initial one.
That is, as the velocity increases in the presence of noise, the 
stress field component $\sigma_{yy}$ fluctuates and  
before $\sigma_{xx}$ becomes 
comparable to $\sigma_{yy}$
the crack ceases be confined to the initial plane. 
In Fig. (\ref{wander}) we plot relative displacements in the $y$ direction for 
two cases. Fig. (\ref{wander}a) shows these displacements just before the crack begins to 
propagate. Fig. (\ref{wander}b) is same plot but just before the first wandering is encountered. 
In the second case the relative displacement in $y$ -direction is 
higher in the row close to the middle one in which crack is propagating.
Recent experiments \cite {bc2} clearly show similar observations. 
There are two velocity thresholds, $v_{c1}$ and $v_{c2}$ such that 
for velocities $v_{c1} \leq v \leq v_{c2}$ surface roughens with almost 
negligible branching while above $v_{c2}$ it branches. The value of 
$v_{c2}$ is around 0.67$c_{R}$, (the Yoffe velocity) where $\sigma_{xx}$ 
becomes comparable to $\sigma_{yy}$.   

\begin{figure} 
\epsfxsize=\hsize \epsfysize = 3.5in \centerline{\epsfbox{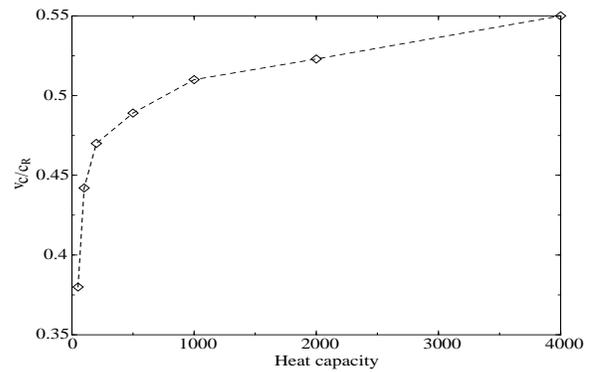}}
\caption{Critical velocity as a function of heat capacity. For large
heat capacity local heating in the vicinity of the tip is 
suppressed.}\label{vvsc}  
\end{figure}

In Fig. (\ref{vvsc}) we show $v_c$ as a function of the heat capacity
of the heat bath. When $C$ is large the tip of the crack remains at
the ambient temperature. We see that about half of the reduction of
$v_c$ is due to tip heating for our parameters.                                                                 
In this paper we have used a naive model to treat fracture in
SiO$_2$. In fact, Griffith used exactly the same sort of considerations 
in his pioneering work on glass \cite{Lawn}. We think
that the qualitative results of this study will survive a more detailed
look at the theory, and that the effect of thermal fluctuations on
$v_c$ is potentially quite large.
The real test for our ideas must come from 
experiment. For example we could imagine looking at the branching
threshold systematically as a function of temperature, or, in materials
like PMMA where the polymer chains vary in length, as a function of
heat capacity. 

We would like to acknowledge useful conversations with O. Pla, F. Guinea,
E. Louis, J. Fineberg, M. Ben-Amar, and M. Adda-Bedia. Supported by
NSF grant DMR 94-20335. S. Ghaisas thanks the University of Michigan
for hospitality.

\end{document}